\documentclass{aa}

\usepackage{natbib}
\bibpunct{(}{)}{;}{a}{}{,} % to follow the A&A style
\usepackage{graphics}   % standard graphics specifications
\usepackage{graphicx}   % alternative graphics specificati.jpgons
\usepackage{epstopdf}
\usepackage[varg]{txfonts}
\usepackage{pdflscape}
\usepackage{hyperref}   %link
\usepackage[flushleft]{threeparttable}
\usepackage{longtable}
\usepackage{caption}
\usepackage{balance}
\hypersetup{
    bookmarks=true,         % show bookmarks bar?
    unicode=false,          % non-Latin characters in Acrobat?s bookmarks
    colorlinks=true,       % false: boxed links; true: colored links
    linkcolor=blue,          % color of internal links (change box color with linkbordercolor)
    citecolor=blue,        % color of links to bibliography
    filecolor=blue,      % color of file links
    urlcolor=blue           % color of external links
}

\begin{document} 

   \title{A strong neutron burst in jet-like supernovae of spinstars}

   \author{
   Arthur Choplin\inst{1,2}, Nozomu Tominaga\inst{1,3} and Bradley S. Meyer\inst{4}
                    }

 \authorrunning{Authors}

      \institute{
Department of Physics, Faculty of Science and Engineering, Konan University, 8-9-1 Okamoto, Kobe, Hyogo 658-8501, Japan
e-mail: arthur.choplin@konan-u.ac.jp
\and Geneva Observatory, University of Geneva, Maillettes 51, CH-1290 Sauverny, Switzerland
\and Kavli Institute for the Physics and Mathematics of the Universe (WPI), The University of Tokyo, 5-1-5 Kashiwanoha, Kashiwa, Chiba, 277-8583, Japan
\and Department of Physics and Astronomy, Clemson University, Clemson, SC 29634-0978, USA
                }
  
   \date{Received / Accepted}

  \abstract
   {Some metal-poor stars have abundance patterns, which are midway between the slow (s) and rapid (r) neutron capture processes.}
   {We show that the helium shell of a fast rotating massive star experiencing a jet-like explosion undergoes two efficient neutron capture processes: one during stellar evolution and one during the explosion. It eventually provides a material whose chemical composition is midway between the s- and r-process. 
}
   {A low metallicity 40~$M_{\odot}$ model with an initial rotational velocity of $\sim 700$~km~s$^{-1}$ was computed from birth to pre-supernova with an extended nuclear network following the slow neutron capture process. 
   A two-dimensional hydrodynamic relativistic code was used to model a $E = 10^{52}$~erg relativistic jet-like explosion hitting the stellar mantle. 
   The jet-induced nucleosynthesis was calculated in post-processing with an optimised network of 1812 nuclei.}
   {During the star's life, heavy elements from $30 \lesssim Z \lesssim 82$ are produced thanks to an efficient s-process, which is boosted by rotation. 
    At the end of evolution, the helium shell is largely enriched in trans-iron elements and in (unburnt) $^{22}$Ne, whose abundance is $\sim 20$ times higher than in a non-rotating model. 
   During the explosion, the jet heats the helium shell up to $\sim 1.5$ GK. 
   It efficiently activates ($\alpha,n$) reactions, such as $^{22}$Ne($\alpha,n$), and leads to a strong n-process with neutron densities of $\sim 10^{19} - 10^{20}$~cm$^{-3}$ during $0.1$~second. 
   This has the effect of shifting the s-process pattern, which was built during stellar evolution, towards heavier elements 
 (e.g. Eu). 
The resulting chemical pattern is consistent with the abundances of the carbon-enhanced metal-poor r/s star CS29528-028, provided the ejecta of the jet model is not homogeneously mixed. 
 }
{
The helium burning zones of rotating massive stars experiences an efficient s-process during the evolution followed by an efficient n-process during a jet-like explosion. 
This is a new astrophysical site which can explain at least some of the metal-poor stars showing abundance patterns midway between the s- and r-process. 
}

\titlerunning{Title}
\authorrunning{Authors}

   \maketitle

\section{Introduction}\label{intro}

Understanding the origin of the elements is among the very topical challenges of modern astrophysics \citep[e.g. the reviews of][]{meyer94, arnould07, thielemann17, cowan19, arnould20}. 
The solar abundances first revealed the existence of a slow (s) and a rapid (r) neutron capture process \citep[][]{suess56, burbidge57}. 
The s-process operates during the life of asymptotic giant branch \citep[AGB, main s-process, e.g.][]{busso99, herwig05, lugaro12, karakas14} and massive stars \citep[weak s-process, e.g.][]{langer89, prantzos90, kappeler11}. Promising sites for the r-process include neutron star mergers \citep[e.g.][]{wanajo14, thielemann17} as well as magneto-rotational supernovae and collapsars from massive stars \citep[e.g.][]{winteler12, nishimura15, halevi18, siegel19}.

Besides the Sun, metal-poor (MP) stars \citep[e.g. the review of][]{beers05, frebel15} are prime targets to investigate the origin of elements since they likely formed from a material that was enriched by just one or a few previous sources. Thus their abundances only reflect one or a few previous nucleosynthetic events. 
A zoo of MP stars was progressively constituted based on the chemical composition of these objects. 
Numerous MP stars have supersolar [C/Fe] ratios and are called carbon-enhanced metal-poor (CEMP) stars. 
At [Fe/H] $\lesssim -3$, most MP stars are CEMP stars and generally do not show clear enhancements in heavy elements \citep[CEMP-no stars, e.g.][]{yong13, norris13, bonifacio15, yoon16}. 
In addition to carbon, most CEMP-no stars have supersolar [N/Fe], [O/Fe], [Na/Fe], [Mg/Fe], and [Al/Fe] ratios. The most discussed scenarios to explain these nearly pristine objects that are enriched in light elements rely on massive stellar models experiencing significant mixing at the time of the supernova and possibly strong fallback \citep[e.g.][]{umeda03, limongi03, nomoto03, iwamoto05, heger10, tominaga07b, tominaga14, ishigaki18} and models experiencing mid to fast rotation during their lives \citep[e.g.][]{meynet06, meynet10, hirschi07, joggerst10, maeder15a, takahashi14, choplin17a, choplin19}. 

At [Fe/H] $\gtrsim -3$, many CEMP stars with clear enhancement of heavy elements were found. This includes CEMP stars with clear s-process signatures \citep[CEMP-s stars, e.g.][]{burris00,simmerer04,sivarani04,placco13} and r-process signatures \citep[CEMP-r stars, e.g.][]{sneden03, ji16,hansen18b,sakari18}. 
The origin of the CEMP-r stars is tightly linked to the r-process site (cf. first paragraph). 
Most of the CEMP-s stars are found in binary systems \citep{hansen16a} and are thought to have acquired their s-processed material, from a now extinct AGB companion, during a (wind) mass transfer episode \citep[e.g.][]{stancliffe08, bisterzo10, masseron10, lugaro12, hollek15, abate15b}. 
An efficient s-process is also expected in rotating massive stars (cf. Sect.~\ref{scenar}) and therefore these objects could explain at least a part of the CEMP-s sample \citep{choplin17letter, banerjee19}. 
In addition to CEMP-s and CEMP-r stars, one also finds 
r+s CEMP stars, which are likely made of a combination of s- and
r-process patterns  \citep[][]{gull18}, and CEMP-r/s stars \citep[also named CEMP-i stars, e.g.][]{roederer16}. This last class shows intermediate abundance patterns for which neither an s-process nor an r-process pattern, and likely nor a combination of both, can reasonably reproduce the abundances \citep{jonsell06,lugaro12,bisterzo12,abate16}. 
An intermediate neutron capture process, that is, an i-process operating
at neutron densities in between the s- and r-process \citep[first named by][]{cowan77}, may be a solution to explain these stars\footnote{Such a process has also been proposed to explain the isotopic signature of presolar grains \citep{jadhav13,liu14}.} \citep{dardelet14, roederer16, hampel16,hampel19}. 
The i-process may operate in AGB or super-AGB stars \citep{herwig11, jones16} in rapidly accreting white dwarfs \citep{denissenkov17, denissenkov19} or in the helium shell of very low (or zero) metallicity massive stars experiencing proton ingestion events during late evolutionary stages \citep{banerjee18, clarkson18}.

   \begin{figure}
   \centering
       \includegraphics[scale=0.47]{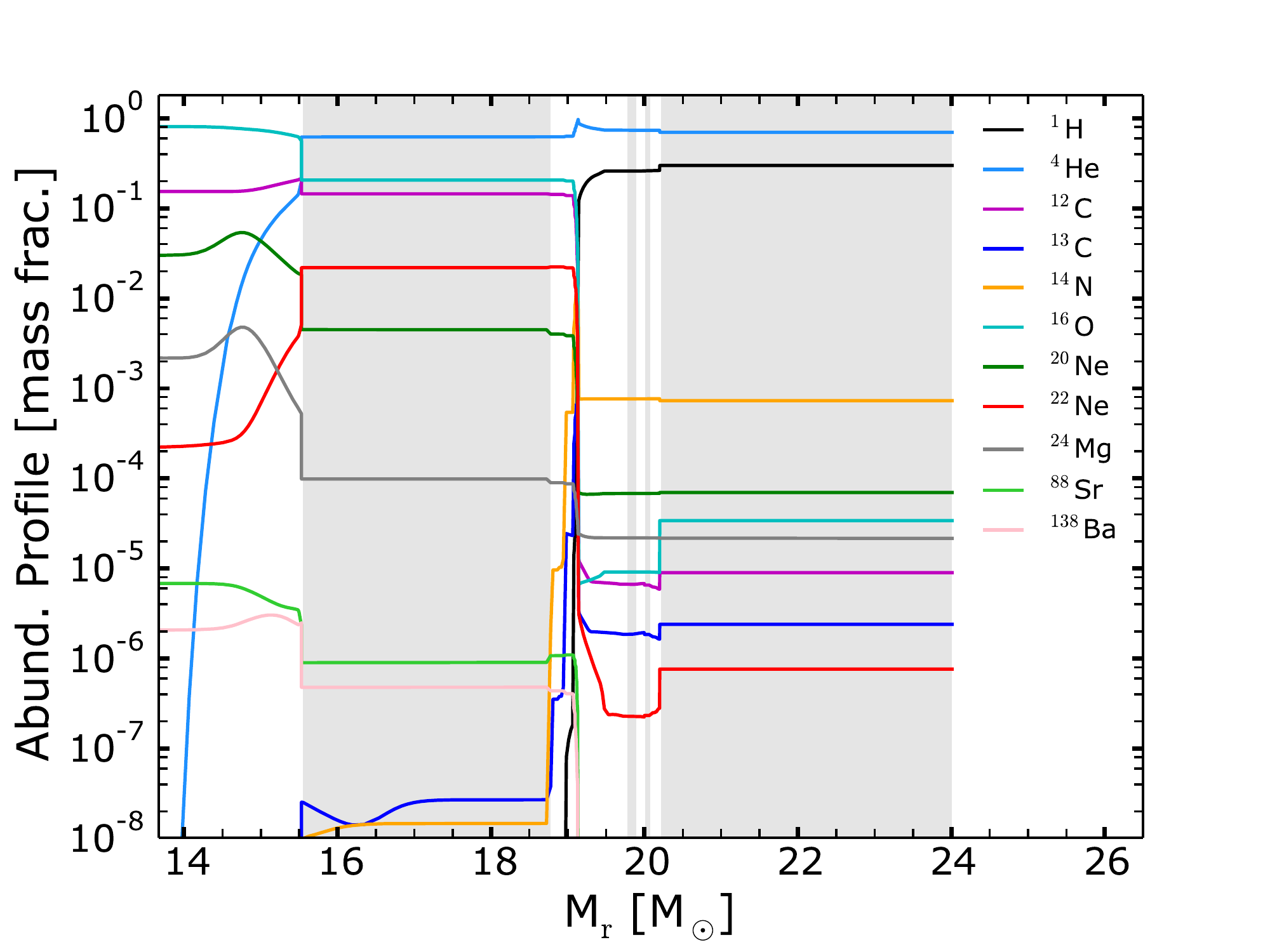}
   \caption{Pre-supernova abundance profile of the rotating 40~$M_{\odot}$
   progenitor model. The outer layers are shown. The shaded areas show the convective zones. 
   }
\label{abprof}
    \end{figure}

   \begin{figure}
   \centering
       \includegraphics[scale=0.46]{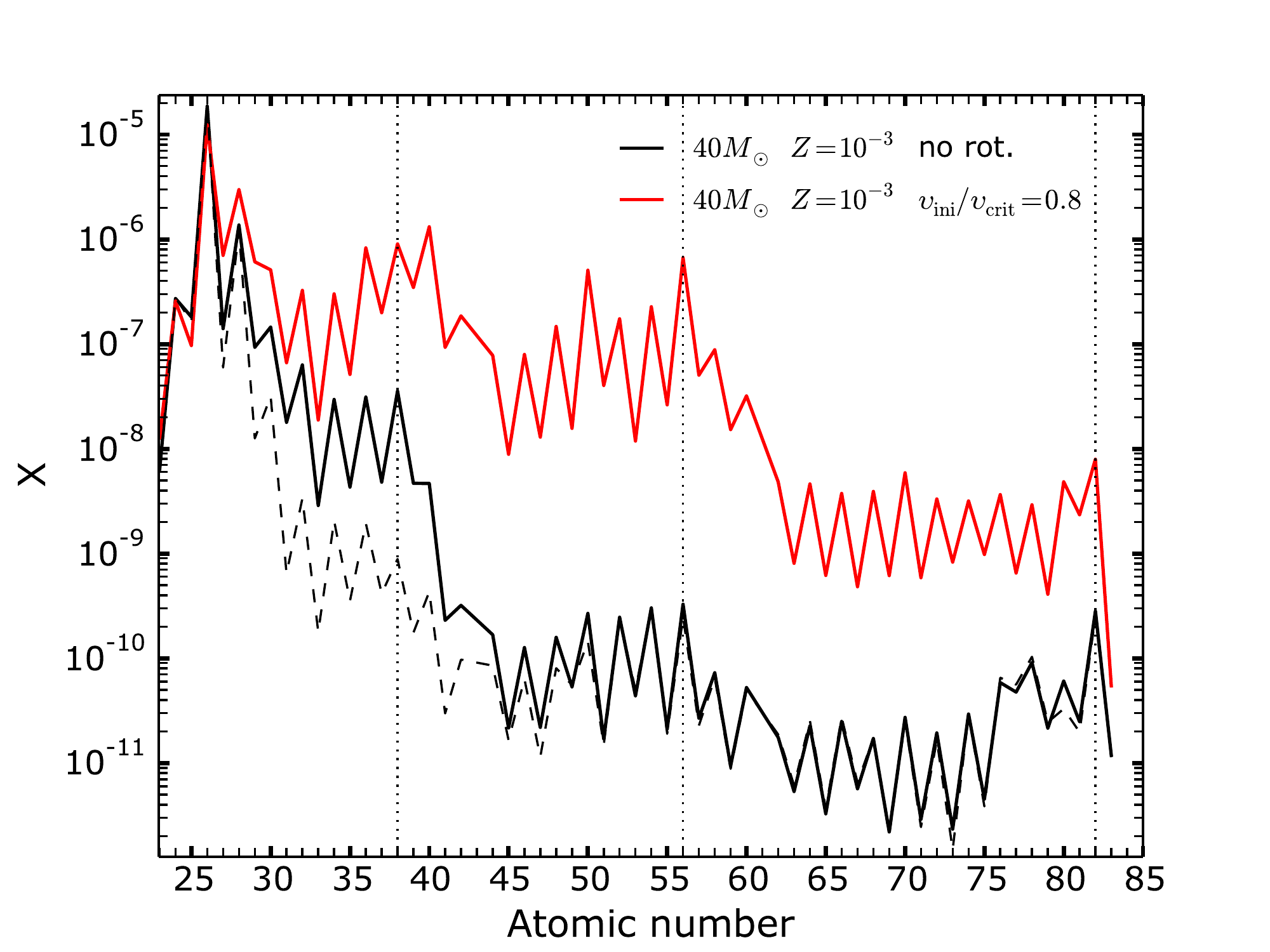}
   \caption{
   Integrated mass fractions of the material above the bottom of the helium shell (after beta decays) at the pre-supernova stage.
   The red (black) solid line shows a low metallicity 40~$M_{\odot}$ model computed with (without) rotation. 
   The black dashed line shows the initial chemical composition of the models. The vertical dotted lines show the location of Sr, Ba, and Pb.
  }
\label{presn}
    \end{figure}

In addition to the s-, r-, and i-processes, one finds the n-process. 
It consists in a short neutron burst in the helium shell of massive stars when the supernova shock wave passes through it \citep{blake76, truran78, thielemann79, blake81, hillebrandt81, rauscher02,  meyer04}. Most of the neutron are provided by the $^{22}$Ne($\alpha$,n) reaction. 
It has been shown that the n-process could explain some anomalous isotopic signatures in meteorites \citep{meyer00, pignatari15,pignatari18}.

Here, we propose a new scenario providing abundance patterns midway between the s- and r-process and which can possibly explain the metal-poor stars having these types of intermediate abundance patterns. 
This scenario relies on two successive neutron capture processes in the same source: an efficient s-process followed by a strong n-process. 
The basic working of it is explained in Sect.~\ref{scenar}. 
The codes used and methods are presented in Sect.~\ref{meth}, the results and comparisons with the metal-poor star C29528-028 are presented in Sect.~\ref{res} and \ref{fitcs}, respectively, followed by discussions in Sect.~\ref{disc}. A summary and the main conclusions are given in Sect.~\ref{concl}.

\section{The helium shell of a rotating massive star hit by a jet}\label{scenar}

This section describes the basic working of the scenario discussed throughout this paper.
By progressively transporting chemical elements from a burning region to another rotational mixing impacts the nucleosynthesis during stellar evolution 
\citep[e.g.][]{meynet00b, heger00, brott11, chieffi13, choplin16}. 
Especially, rotation has been found to increase the production of $^{13}$C and $^{22}$Ne, which release neutrons by ($\alpha, n$) reactions. 
It boosts the efficiency of the s-process in massive stars, mainly during the core helium burning stage 
\citep[][see also Sect.~\ref{rotind}]{pignatari08,frischknecht12,frischknecht16,limongi18,choplin18,banerjee19}.

   \begin{figure*}
   \centering
       \includegraphics[scale=0.49, trim = 0cm 0cm 0cm 0cm]{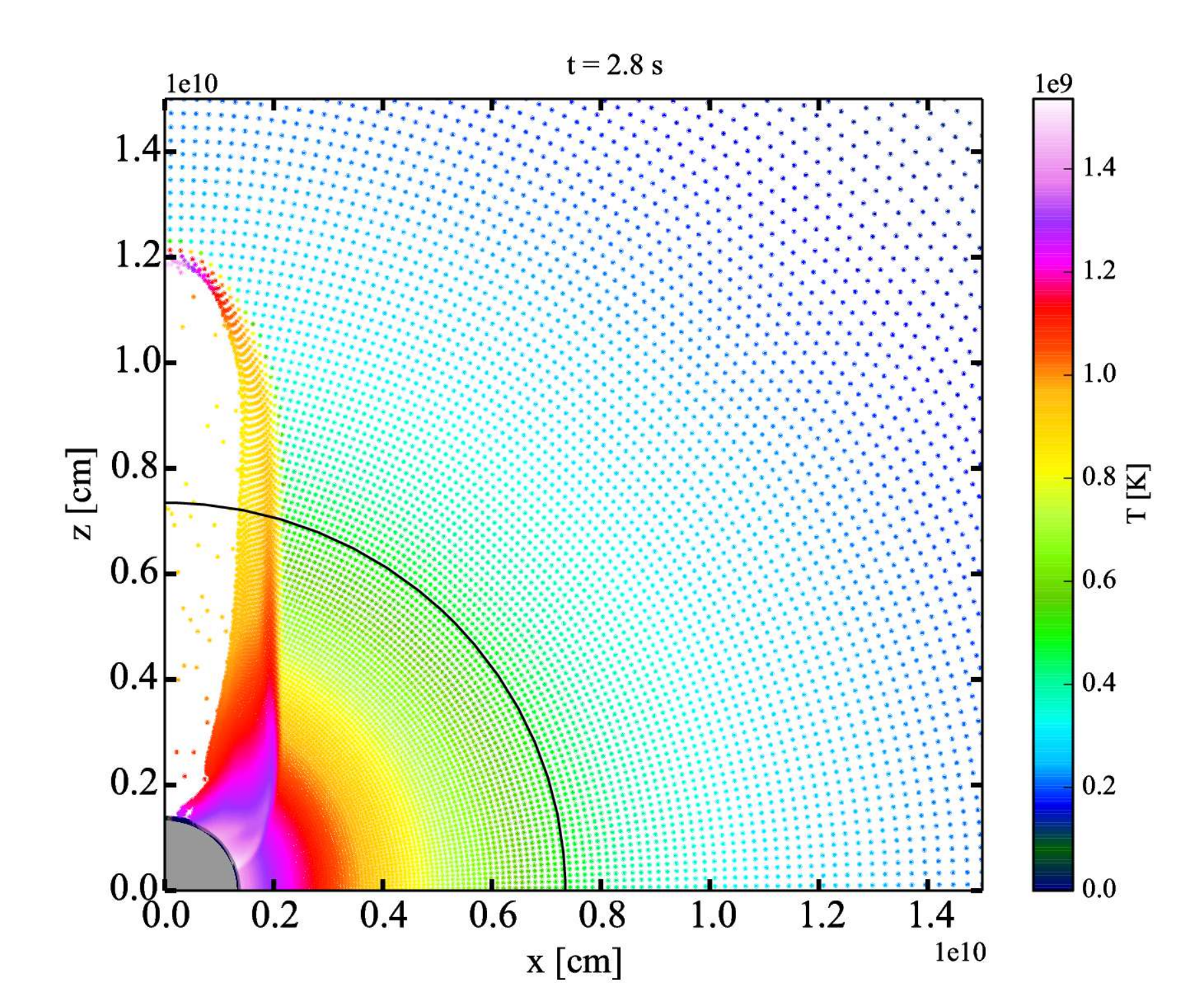}
      \includegraphics[scale=0.49, trim = 0cm 0cm 0cm 0cm]{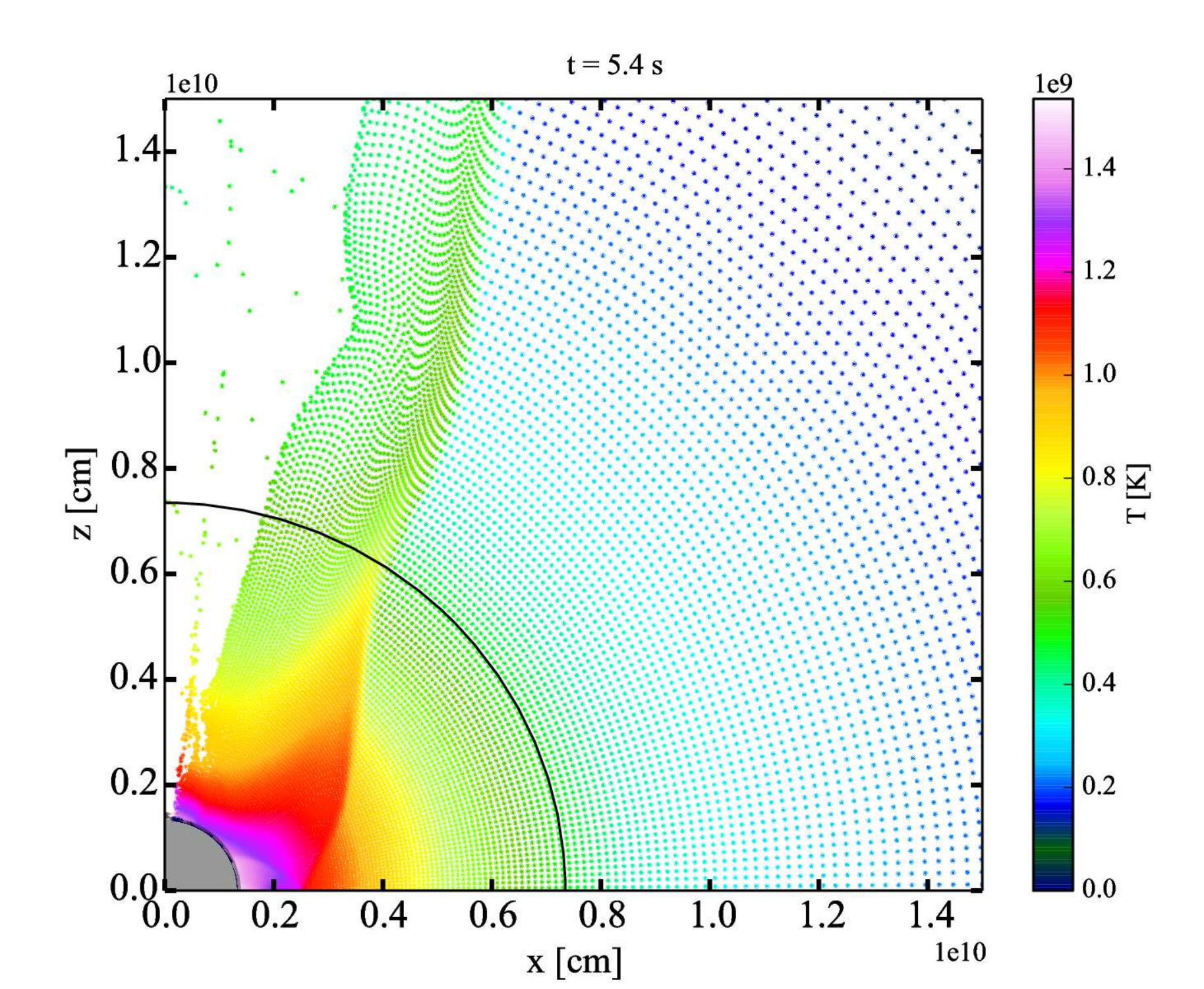}
   \caption{Location of the mass particles after 2.8 s (left panel) and 5.4 s (right panel). The colour shows the temperature of the mass particles. The inner grey circle shows the central remnant. The black circle with a radius of $\sim 0.7 \times 10^{10}$~cm shows the location of the bottom of the helium shell. 
   }
\label{jet}
    \end{figure*}

As mentioned in Sect.~\ref{intro}, at the time of the supernova, the shock wave passing through the helium shell can trigger the n-process.
Its efficiency very much depends on the pre-supernova amount of $^{22}$Ne and the ability of the explosion to burn it during the shock passage. 
In case of a rotating progenitor, the n-process can be particularly efficient. On the one hand, rotation produces more $^{22}$Ne during the evolution and allows for a higher amount of unburnt $^{22}$Ne in the helium shell at the pre-supernova stage. On the other hand, rotation may trigger energetic jet-like explosions \citep[e.g.][]{woosley93, macfadyen99,woosley06}, thus leading to high temperatures in the helium shell and therefore an efficient activation of the $^{22}$Ne$(\alpha,n)$ reaction.

The combined effect of fast rotation and jet-like explosion can induce an efficient s-process during evolution and  an efficient n-process during explosion. In this paper we examine the combination of these two effects.

\section{Methods}\label{meth}

We modelled the evolution and explosion of a 40~$M_{\odot}$ star by combining a one-dimensional stellar evolution code with a two-dimensional explosion code. 
The Geneva stellar evolution code \citep{eggenberger08} was used to compute a pre-supernova 40 $M_{\odot}$ model at a metallicity of $Z=10^{-3}$ with an initial rotation rate on the zero-age main sequence of $\upsilon_{\rm ini}/\upsilon_{\rm crit} = 0.8$\footnote{The initial equatorial velocity is $\upsilon_{\rm ini}$ and $\upsilon_{\rm crit}$ is the initial equatorial velocity at which the gravitational acceleration is balanced by the centrifugal force. It is defined as $\upsilon_{\rm crit} = \sqrt{ \frac{2GM}{3R_{\rm pb}}}$ , where $R_{\rm pb}$ is the the polar radius at the break-up  velocity \citep[see][]{maeder00a}.}. 
This corresponds to an initial equatorial velocity of 704~km~s$^{-1}$. 
The nuclear network was coupled to the evolution and used throughout the entire process. It comprises 737 isotopes from hydrogen to polonium. 
We refer to \cite{choplin20} for more details on the initial set up of the model.

The jet-like explosion was modelled with a two-dimensional relativistic hydrodynamical code \citep{tominaga07a, tominaga09}. 
At $t=0$, a jet was launched at the mass coordinate 1.93 $M_{\odot}$, lasting for 10~seconds with a constant energy deposition rate of $10^{51}$ erg~s$^{-1}$. The total energy deposited is then $10^{52}$~erg, which corresponds to the typical energy of a hypernova \citep[e.g.][]{nomoto04,nomoto06}. 
The opening angle is $5$~degrees and the Lorentz factor is $\Gamma  = 100$. 
The hydrodynamics is followed for $5 \times 10^5$~seconds.
The thermodynamic histories are recorded by $2.5 \times 10^4$ mass particles representing Lagrangian mass elements of the stellar mantle. The initial grid is linear with a resolution of 100 along the $\theta-$direction and logarithmic with a resolution of 250 in the $r-$direction.
The nucleosynthesis of mass particles initially located in the helium shell was calculated in post-processing with a neutron-rich nuclear network of 1812 isotopes.
Nucleosynthesis tests with different network sizes were carried out in order to reduce the network size as much as possible without missing isotopes.

   \begin{figure}
   \centering
       \includegraphics[scale=0.44, trim=1cm 0cm 0cm 0cm]{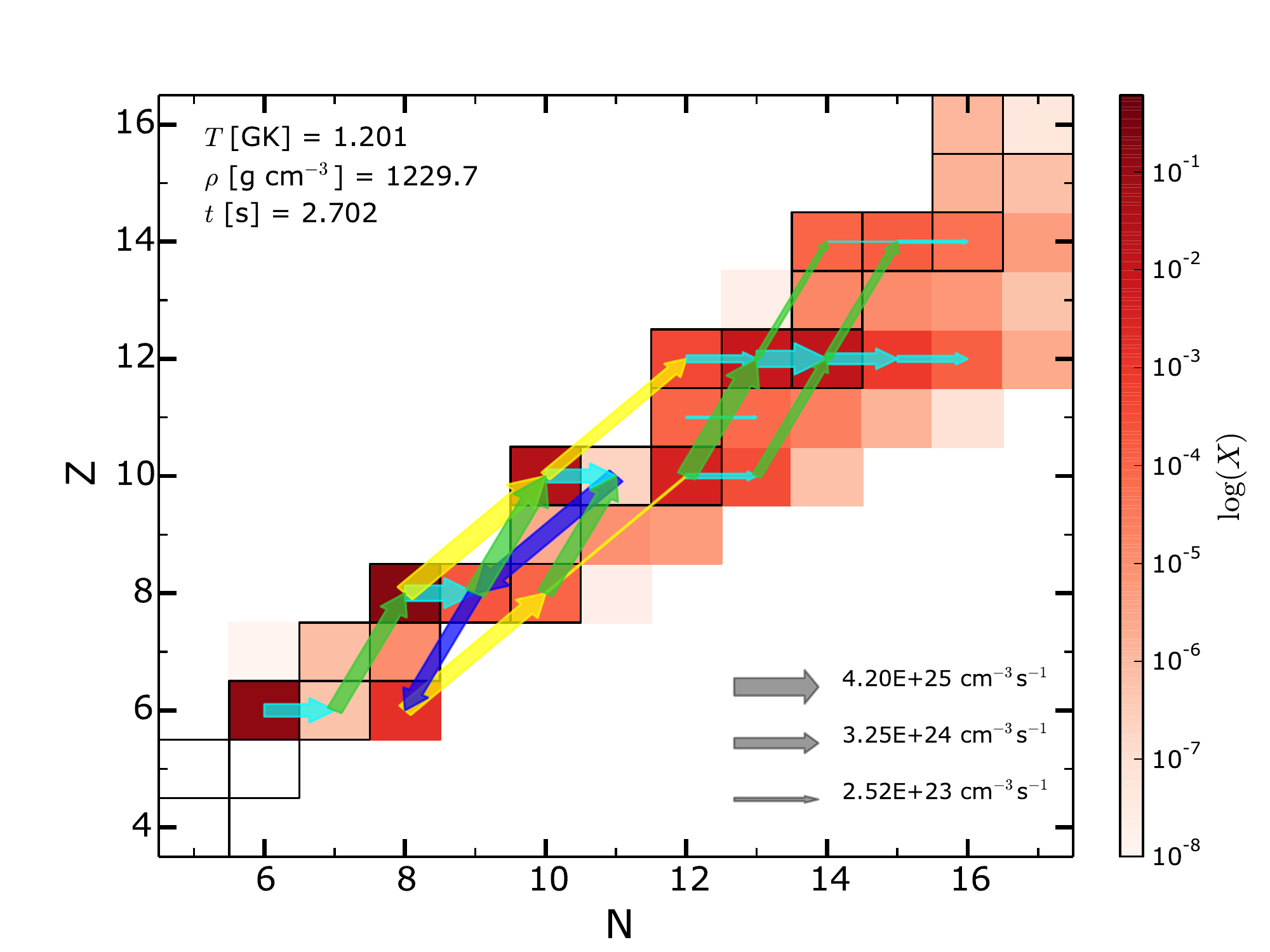}
   \caption{Flowchart of a mass particle with $M_{\rm r,ini} = 16.53$~$M_{\odot}$ 
   (i.e. originally in the helium shell, cf. Fig.~\ref{abprof}) 
  and $\theta_{\rm ini} = 3.15^{\circ}$ at its temperature peak. 
  Mass fractions of isotopes are shown by the red colourmap. Black squares denote stable isotopes. Net reaction rates, considering forward and reverse reactions, are shown by arrows. 
  Arrows are coloured as a function of the reaction type: yellow for $(\alpha,\gamma)$, cyan for $(n,\gamma)$, green for $(\alpha,n),$ and blue for $(\gamma,\alpha)$ and $(n,\alpha)$. 
  Rates below $7 \times 10^{22}$~cm~s$^{-1}$ are not shown.}
\label{flow1}
    \end{figure}

\section{Results}\label{res}

During its life, this rotating 40~$M_{\odot}$ model loses 16~$M_{\odot}$ through stellar winds: 
8~\% of the mass is lost during the main sequence, 82~\% during core helium burning, and 10~\% after core helium burning.
The final mass of the model is then $M_{\rm fin} = 24$~$M_{\odot}$ with a hydrogen envelope that is mainly composed of helium (71~\% in mass). The final surface hydrogen mass fraction is 0.3 (Fig.~\ref{abprof}) and the total hydrogen mass left in the star is 1.4~$M_{\odot}$. 
For comparison purposes, the same model without rotation loses 6.42~$M_{\odot}$, has a final hydrogen envelope of 18.29~$M_{\odot}$ (composed of $\sim 45$~\% hydrogen and $\sim 55$~\% helium), has a final surface hydrogen mass fraction of 0.73, and the total hydrogen mass that is left in the star is 8.2~$M_{\odot}$.

\subsection{Rotation-induced nucleosynthesis during stellar evolution}\label{rotind}

We recall below how rotation boosts the s-process during the evolution of massive stars \citep[cf. discussions in][]{pignatari08, frischknecht16,limongi18,choplin18,banerjee19}. 
During the core helium burning phase, $^{12}$C and $^{16}$O are transported by rotation-induced mixing from the helium core to the hydrogen shell, which boosts the CNO cycle in the hydrogen shell and synthesises primary\footnote{Produced from the initial hydrogen and helium content of the star, but not from the initial metals.} $^{13}$C and $^{14}$N. 
Due to the growth of the convective helium core combined with the backward diffusion of hydrogen burning products, the extra $^{13}$C and $^{14}$N are engulfed in the helium core
The $^{14}$N is quickly transformed into $^{22}$Ne through $^{14}$N($\alpha,\gamma$)$^{18}$F($\beta^+$)$^{18}$O($\alpha,\gamma$)$^{22}$Ne. Neutrons, which are mainly released by $^{13}$C($\alpha,n$) and $^{22}$Ne($\alpha,n$), trigger the s-process provided any heavy seed (e.g. $^{56}$Fe) is present in the helium burning core.

In the present model, this effect boosts the s-process by a factor of $\sim 10 - 1000$ for elements with $35<Z<82$ (Fig.~\ref{presn}). The same model without rotation only experiences a modest enhancement of elements with  $Z<45$ (Fig.~\ref{presn}, black line).
At the pre-supernova stage, the mass fraction of $^{22}$Ne that is left (unburnt) in the helium shell is $2.2 \times 10^{-2}$ (Fig.~\ref{abprof}), which is $\sim 20$ times higher than for the same model but without rotation.

   \begin{figure}
   \centering
       \includegraphics[scale=0.32 , trim = 0.5cm 0cm 0cm 0cm]{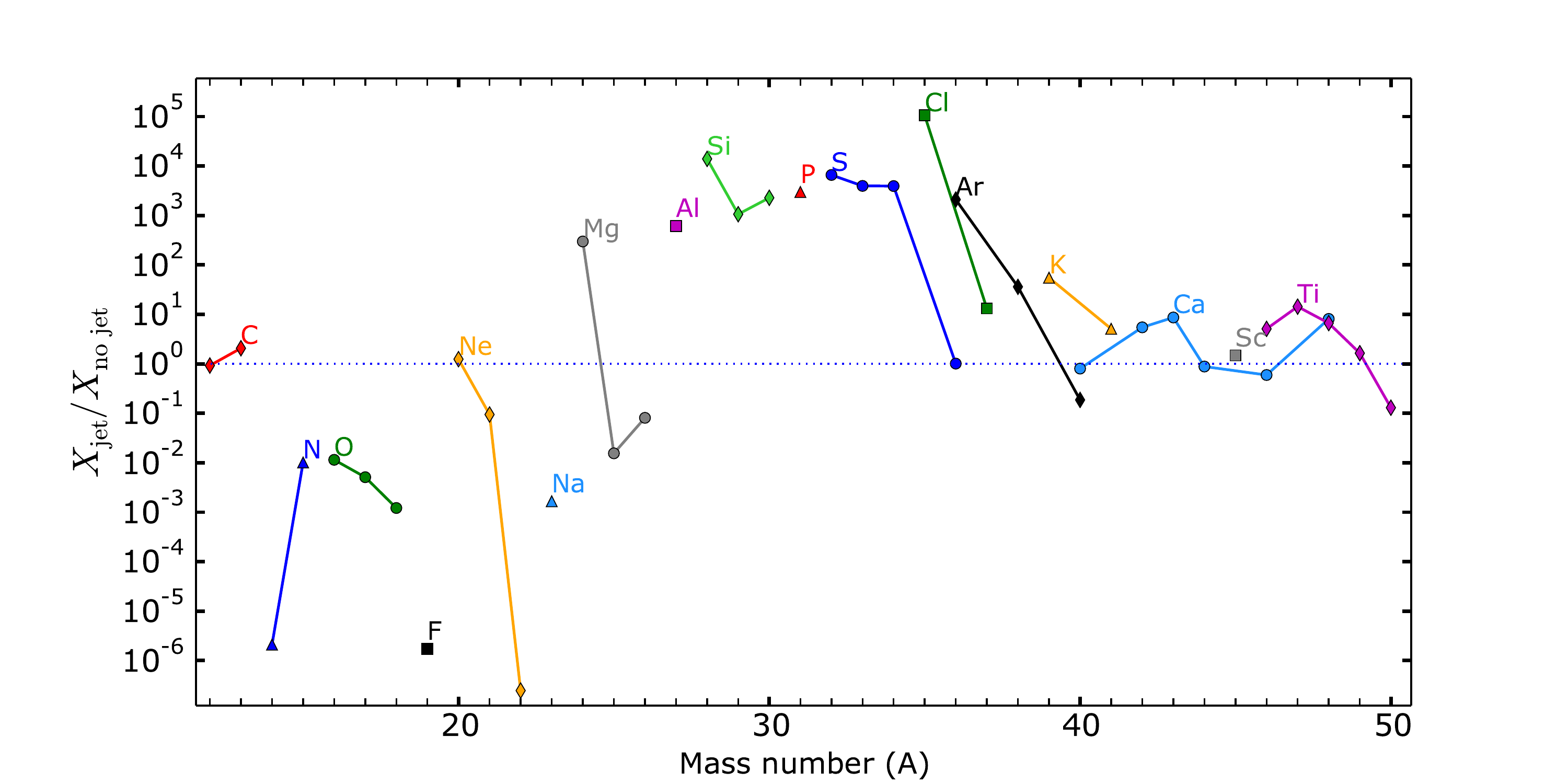}
   \caption{
   Effect of the jet on C$-$Ti elements for a mass particle with $M_{\rm r,ini} = 15.78$~$M_{\odot}$ and $\theta_{\rm ini} = 0.45^{\circ}$. We note that $X_{\rm jet}$ are the post-jet mass fractions and 
   $X_{\rm no \, jet}$ are the mass fractions of the same mass particle that would not have experienced a jet explosion. In both cases, elements are beta decayed.
   }
\label{isochains}
    \end{figure}

   \begin{figure}
   \centering
       \includegraphics[scale=0.33, trim=1cm 0cm 0cm 0cm]{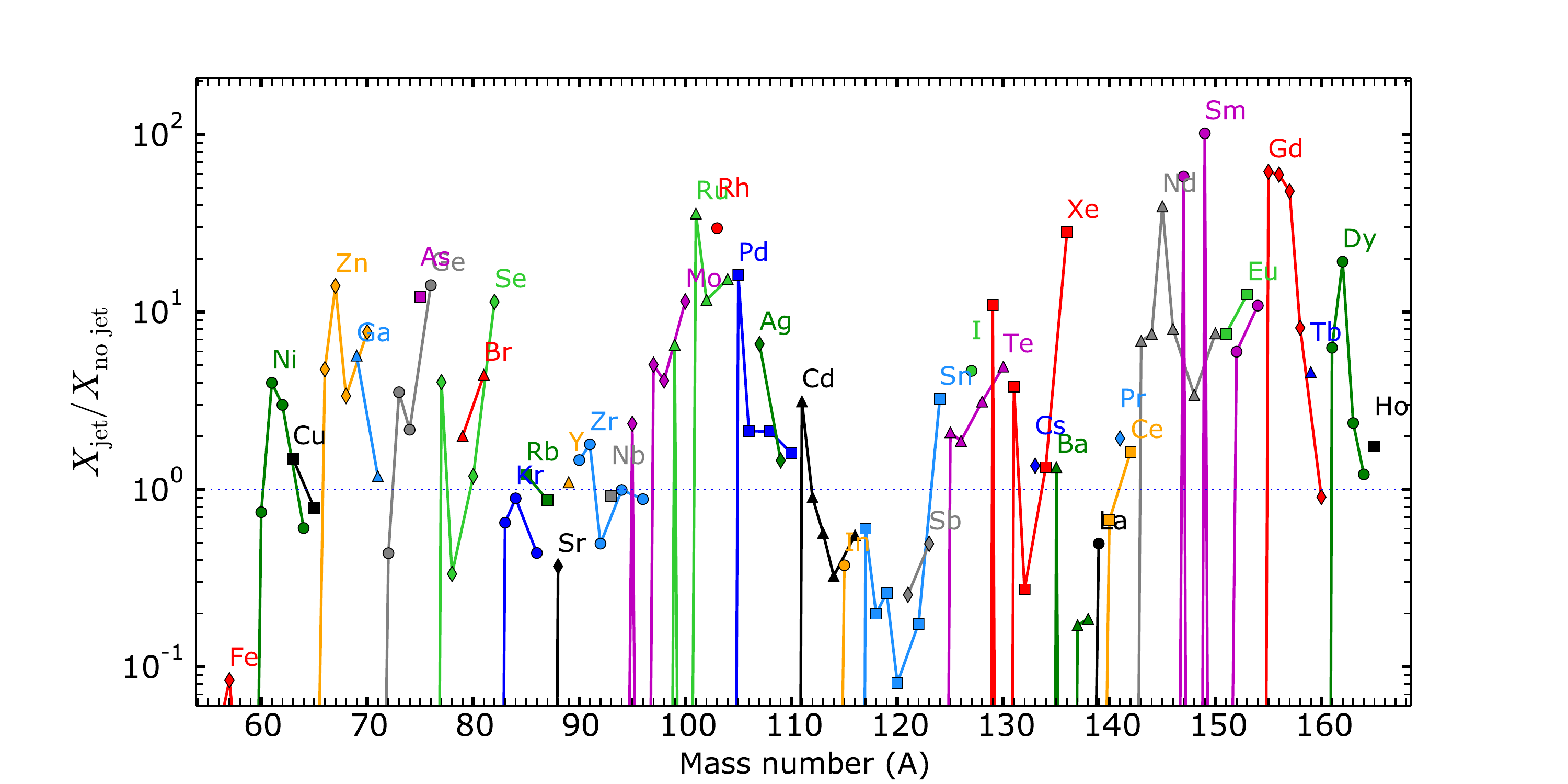}
   \caption{Same as Fig.~\ref{isochains}, but for Fe$-$Ho elements and for a mass particle with $M_{\rm r,ini} = 16.53$~$M_{\odot}$ 
  and $\theta_{\rm ini} = 3.15^{\circ}$ (same mass particle as in Fig.~\ref{flow1}).
   }
\label{isochains2}
    \end{figure}

   \begin{figure}
   \centering
       \includegraphics[scale=0.44]{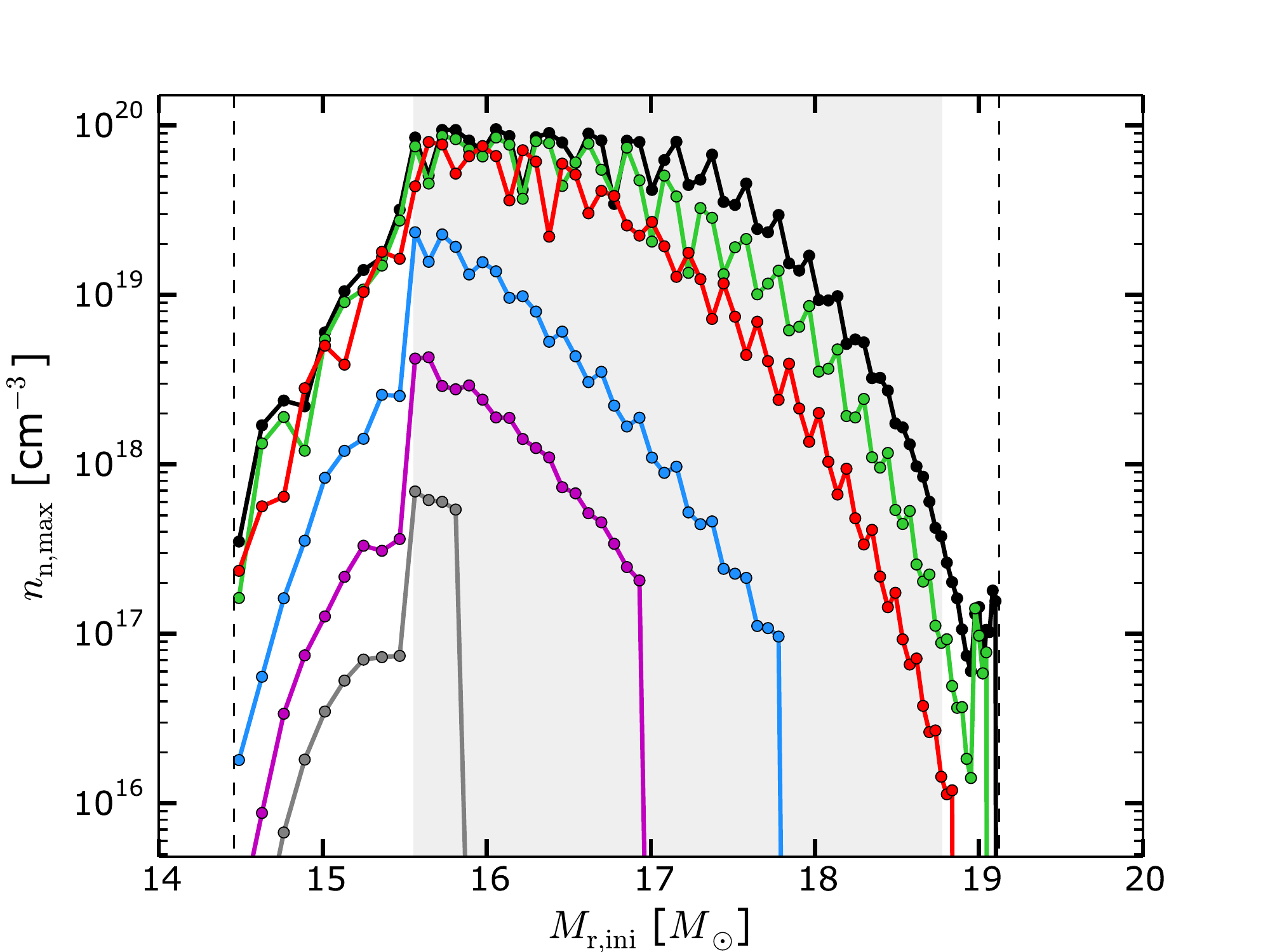}
       \includegraphics[scale=0.44]{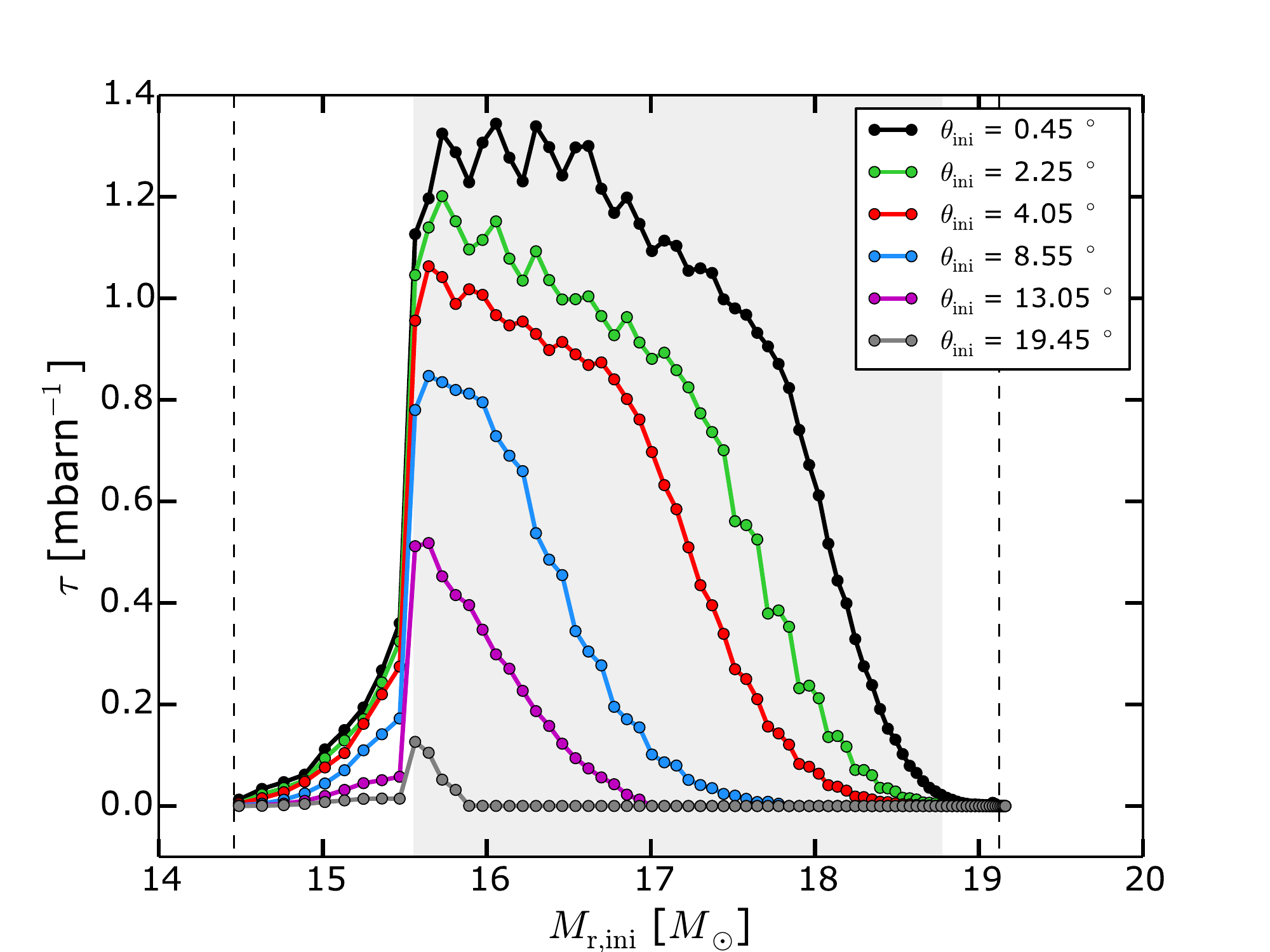}
   \caption{Maximum neutron density (top) and neutron exposure (bottom) during the jet explosion as a function of the initial mass coordinate. 
  Coloured lines show different initial angles. Each point represents a mass particle with initial coordinates $\theta_{\rm ini}$ and $M_{\rm r, ini}$. The two dashed lines show the initial location of the bottom of helium and hydrogen shells. The shaded areas show the convective zone at the pre-jet stage (see also Fig.~\ref{abprof}). 
   }
\label{nexp}
    \end{figure}

\subsection{Jet-induced nucleosynthesis: Light elements}

The jet penetrates into the helium shell after 2 seconds. The mass particles located at the bottom of the helium shell have pre-jet temperatures of $0.3-0.4$ GK. Those with $\theta_{\rm ini} \leq 10^{\circ}$ reach 1~$<$~$T$~$<$~1.5~GK during the passage of the jet (Fig.~\ref{jet}).

Close to the jet axis, the high temperature activates 
$^{22}$Ne($\alpha,n$), but also $^{13}$C($\alpha,n$) and $^{17}$O($\alpha,n$), the isotopes $^{13}$C and $^{17}$O being produced by $^{12}$C($n,\gamma$) and $^{16}$O($n,\gamma$) (Fig.~\ref{flow1}). 
Even closer to the jet axis and deeper in the helium shell, the higher temperature efficiently produces $^{24}$Mg, $^{28}$Si, and $^{32}$S (Fig.~\ref{isochains}) through ($\alpha,\gamma$) reactions. 
Some ($\alpha,p$) reactions, such as $^{24}$Mg($\alpha,p$)$^{27}$Al and $^{27}$Mg($\alpha,p$)$^{30}$Si, also become strong 
and, therefore, they release protons, which activate some ($p,\gamma$) reactions. 
For instance, $^{31}$P is enhanced (Fig.~\ref{isochains}) through $^{30}$Si($p,\gamma$)$^{31}$P.
The high $^{35}$Cl abundance comes from the chain $^{24}$Mg($\alpha,p$)$^{27}$Al($\alpha,p$)$^{30}$Si($\alpha,\gamma$)$^{34}$S($p,\gamma$)$^{35}$Cl.  
The high $^{36}$Ar and $^{39}$K mainly come from $^{34}$S($p,\gamma$)$^{35}$Cl($p,\gamma$)$^{36}$Ar and $^{34}$S($\alpha,\gamma$)$^{38}$Ar($p,\gamma$)$^{39}$K, respectively.
Some Ti is also formed thanks to $^{44}$Ca($p,\gamma$)$^{45}$Sc($p,\gamma$)$^{46}$Ti.

\subsection{Jet-induced nucleosynthesis: Heavy elements}\label{jetheavy}

\begin{figure}
   \centering
       \includegraphics[scale=0.285, trim = 0cm 0cm 0cm 0cm]{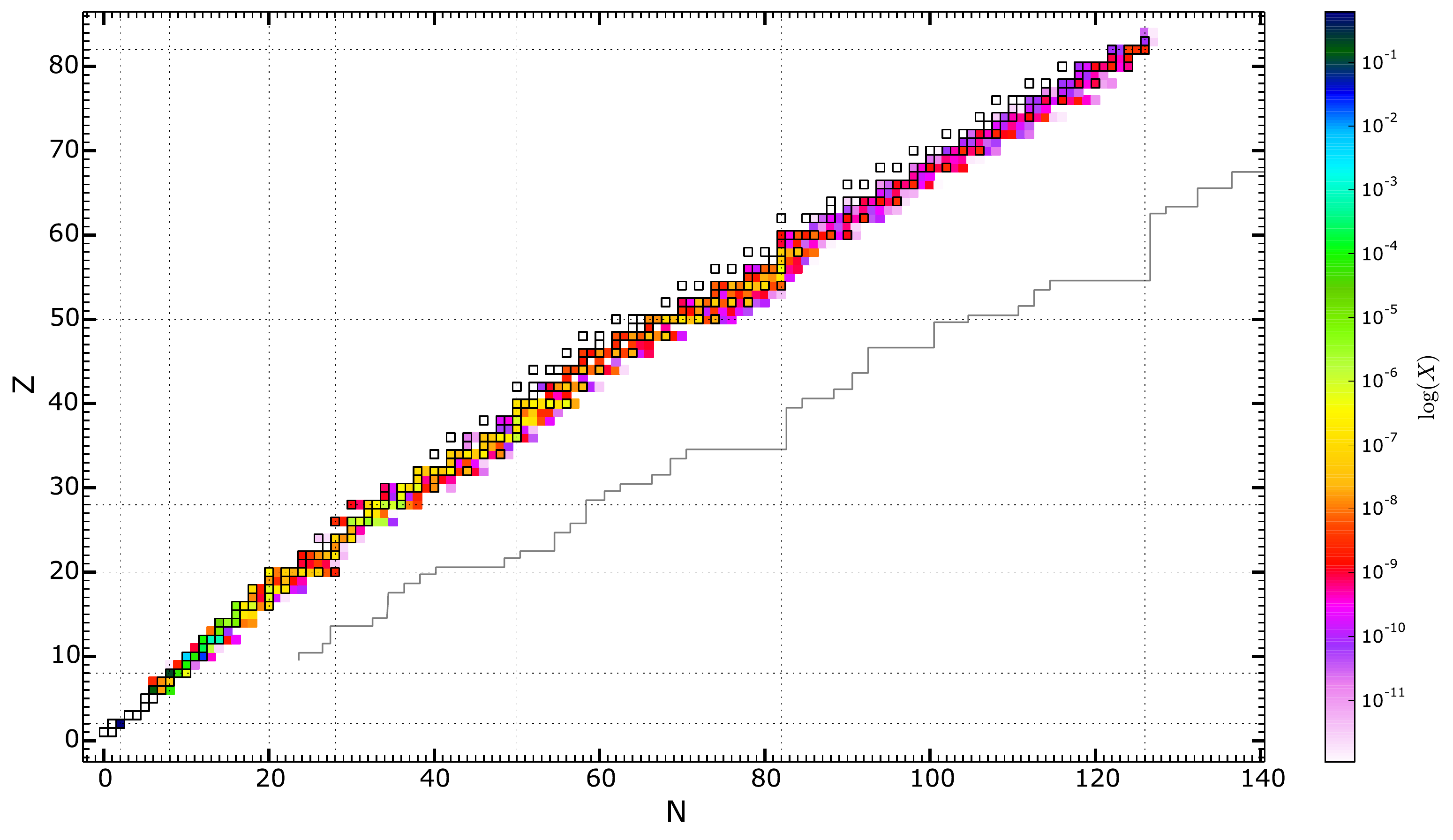}
       \includegraphics[scale=0.285, trim = 0cm 0cm 0cm 0cm]{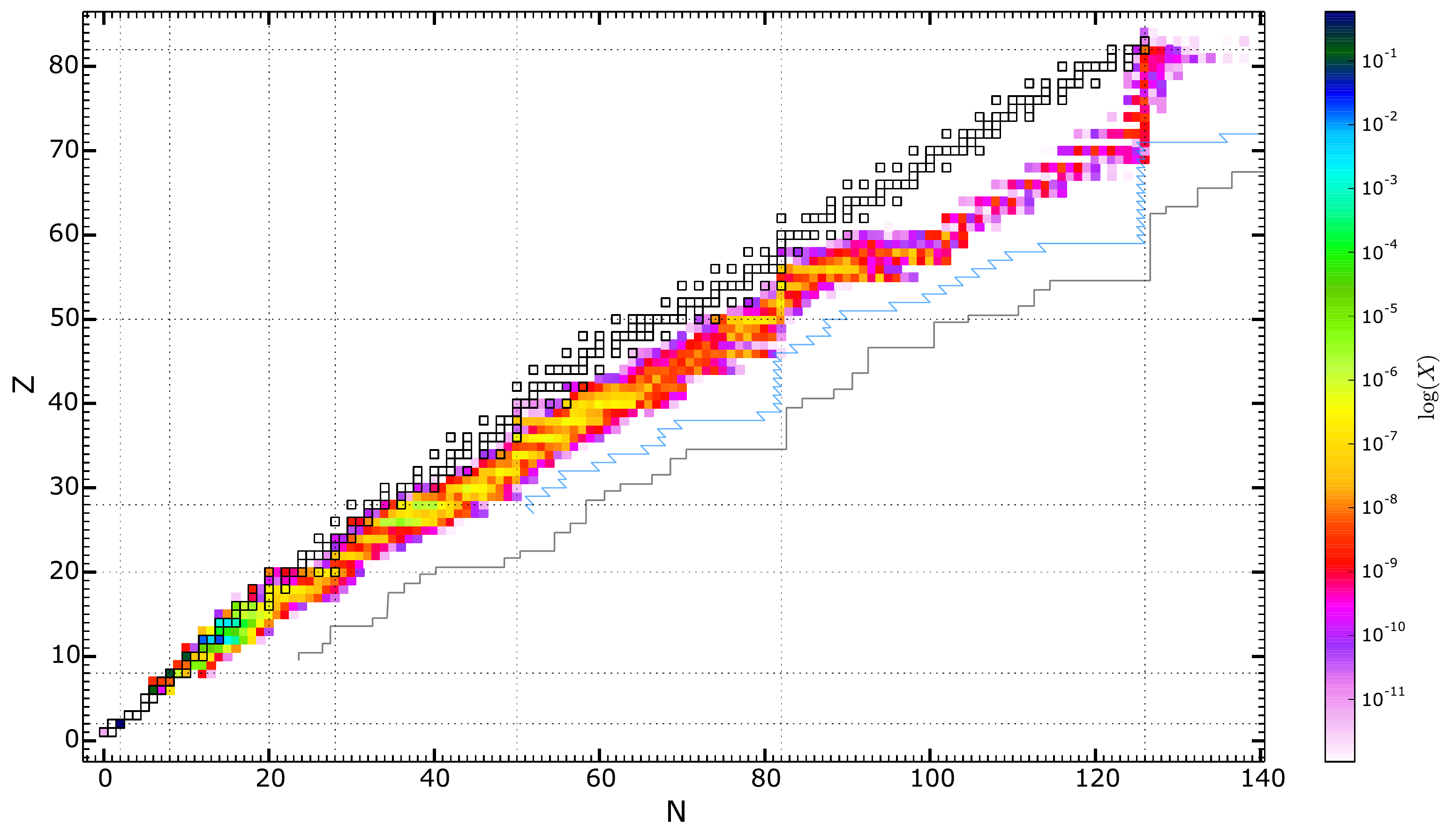}
   \caption{Pre-jet mass fractions (top panel) and post-jet mass fractions (before beta decays, bottom panel) for a mass particle at $M_{\rm r,ini} = 16.53$~$M_{\odot}$ and $\theta_{\rm ini} = 3.15^{\circ}$ (same mass particle as in Fig.~\ref{flow1}). 
   Black squares denote stable isotopes. 
   The light blue line shows a typical r-process flow for comparison purposes \citep[taken from][their Fig.~5]{arnould20}.
   The grey line shows the neutron drip line. 
   Thin dotted lines show the location of the magic numbers. 
   }
\label{nucchart}
    \end{figure}

During the jet explosion, the $(\alpha,n)$ reactions lead to a burst of neutrons. 
For mass particles that are initially along the jet axis ($\theta_{\rm ini} \lesssim 5 ^{\circ}$) and with initial mass coordinates of $M_{\rm r,ini} \sim 15.5-18$~$M_{\odot}$, the neutron density reaches $n_{\rm n} > 10^{19}$~cm$^{-3}$ during $\sim 0.1$~s (Fig.~\ref{nexp}, left panel). 
The neutron exposure\footnote{The neutron exposure was computed as $\tau = \int n_{\rm n} \sqrt{2 k_{\rm B} T / m_{\rm n}} dt$ with $n_{\rm n}$ the neutron density, $k_{\rm B}$ the Boltzmann constant, $T$ the temperature, and $m_{\rm n}$ the neutron mass.} of these mass particles goes up to $1.35$~mbarn$^{-1}$ (Fig.~\ref{nexp}). 
Despite $n_n$ values being close to r-process values, the abundant light elements in the helium shell (especially $^{12}$C and $^{16}$O) act as strong neutron poisons. This reduces the amount of available neutron for heavy seeds and does not lead to r-process nucleosynthesis.
Nevertheless, during the burst, isotopes that are far from the valley of stability are produced (Fig.~\ref{nucchart}, bottom panel). 
The pre-jet abundance pattern is therefore shifted towards higher mass numbers compared to the initial pattern (Fig.~\ref{aba}). 
In particular, the first (A~$\sim 80-90$) and second (A~$\sim 130-140$) s-process peaks that were built during stellar evolution are shifted to higher atomic masses (see also Fig.~\ref{isochains2}). Details on the nucleosynthesis are given below.

During the jet explosion, $^{56,57,58,60}$Fe are mostly transformed into $^{60-64}$Fe by neutron captures. After the explosion, $^{60-64}$Fe eventually decay to the stable $^{60,61,62,64}$Ni and $^{63}$Cu isotopes. 
This has the overall effect of destroying Fe and somewhat increasing Ni and Cu (Fig.~\ref{isochains2} and \ref{abz}).
The $^{59}$Co isotope, which is the only stable isotope of Co, is mainly transformed into $^{62-69}$Co isotopes which decay to Ni, Cu, Zn, and $^{69}$Ga. 
The production of $^{59}$Fe, $^{59}$Mn, and $^{59}$Cr, which will decay to $^{59}$Co, is too small to replenish $^{59}$Co and therefore the final Co abundance stays low (Fig.~\ref{abz}). 
The higher final Zn abundance is mainly due to the decay of $^{66,67,68}$Ni to $^{66,67,68}$Zn isotopes. 
The final Mo is also enhanced, mainly thanks to the decay of neutron-rich Zr ($^{96-100}$Zr) isotopes to stable Mo isotopes. The final Ru is enhanced thanks to the decay of the neutron-rich Zr ($^{101, 102, 104}$Zr), Nb, and Mo isotopes to stable Ru isotopes. We note that $^{103}$Rh, which is the only stable isotope of Rh, is largely enhanced by the decay of $^{103}$Zr, $^{103}$Nb, and $^{103}$Mo.
It is similar for Nd, Sm, Eu, Gd, Tb, and Dy ($Z = 60-66$) elements whose final abundances are raised (Fig.~\ref{isochains2} and \ref{abz}) because of the decays of the neutron-rich isotopes of Cs, Ba, La, Ce, and Pr ($Z = 55-59$).
The final Tm, Lu, and Ta abundances are lower because they experience efficient neutron captures until the neutron magic number 126 (Fig.~\ref{nucchart}), and they are not efficiently replenished by neutron-rich isotopes for elements with a lower atomic mass. For instance, most of the initial $^{169-173}$Tm are transformed into $^{195}$Tm, which eventually decay to $^{195}$Pt, and Tm is only slightly replenished by the decay of $^{169}$Sm, $^{169}$Eu, and $^{169}$Gd, whose post-jet mass fractions are small (about $10^{-11}$). 
The abundance of $^{209}$Bi also increases (Fig.~\ref{abz}) because of the decay of the unstable $^{209}$Tl and $^{209}$Pb isotopes, which are produced during the jet explosion. Furthermore,
Pb is not affected by the jet explosion because of a closed loop of beta and alpha decays leading back to Pb (or possibly $^{209}$Bi). 
For instance, any unstable $^{210}$Pb produced during the explosion experiences double beta decay to $^{210}$Po, and then alpha decay to the stable $^{206}$Pb isotope. 

The stable isotopes, which are shielded by isotopes with the same mass and a lower atomic number, are not replenished by beta decays and are therefore completely destroyed by the jet. 
This is the case for $^{136}$Ba or $^{148}$Sm (Fig.~\ref{isochains2}) whose final mass fractions are nearly zero because they are shielded by the stable $^{136}$Xe and $^{148}$Nd isotopes, respectively.

\subsection{Case of a spherical explosion}\label{sph}

\begin{figure}
   \centering
       \includegraphics[scale=0.325, trim = 1cm 0cm 0cm 0cm]{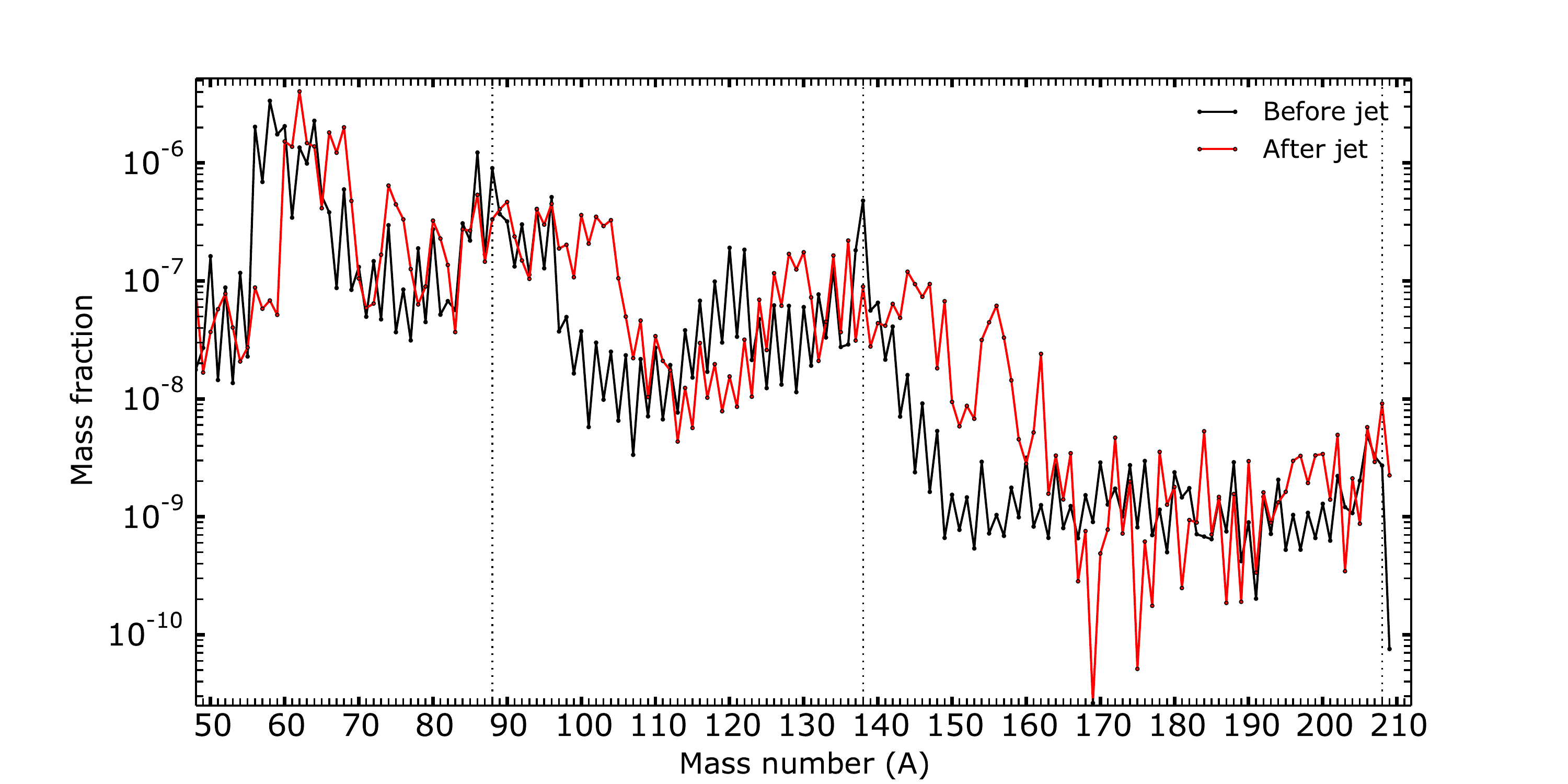}
   \caption{Pre-jet and post-jet mass fractions for the mass particle at $M_{\rm r,ini} = 16.53$~$M_{\odot}$ and $\theta_{\rm ini} = 3.15^{\circ}$ (the same mass particle as in Fig.~\ref{abprof}, \ref{flow1}, and \ref{nucchart}). The vertical dotted lines show the location of $^{88}$Sr, $^{138}$Ba, and $^{208}$Pb. 
   }
\label{aba}
    \end{figure}

\begin{figure}
   \centering
       \includegraphics[scale=0.325, trim = 1cm 0cm 0cm 0cm]{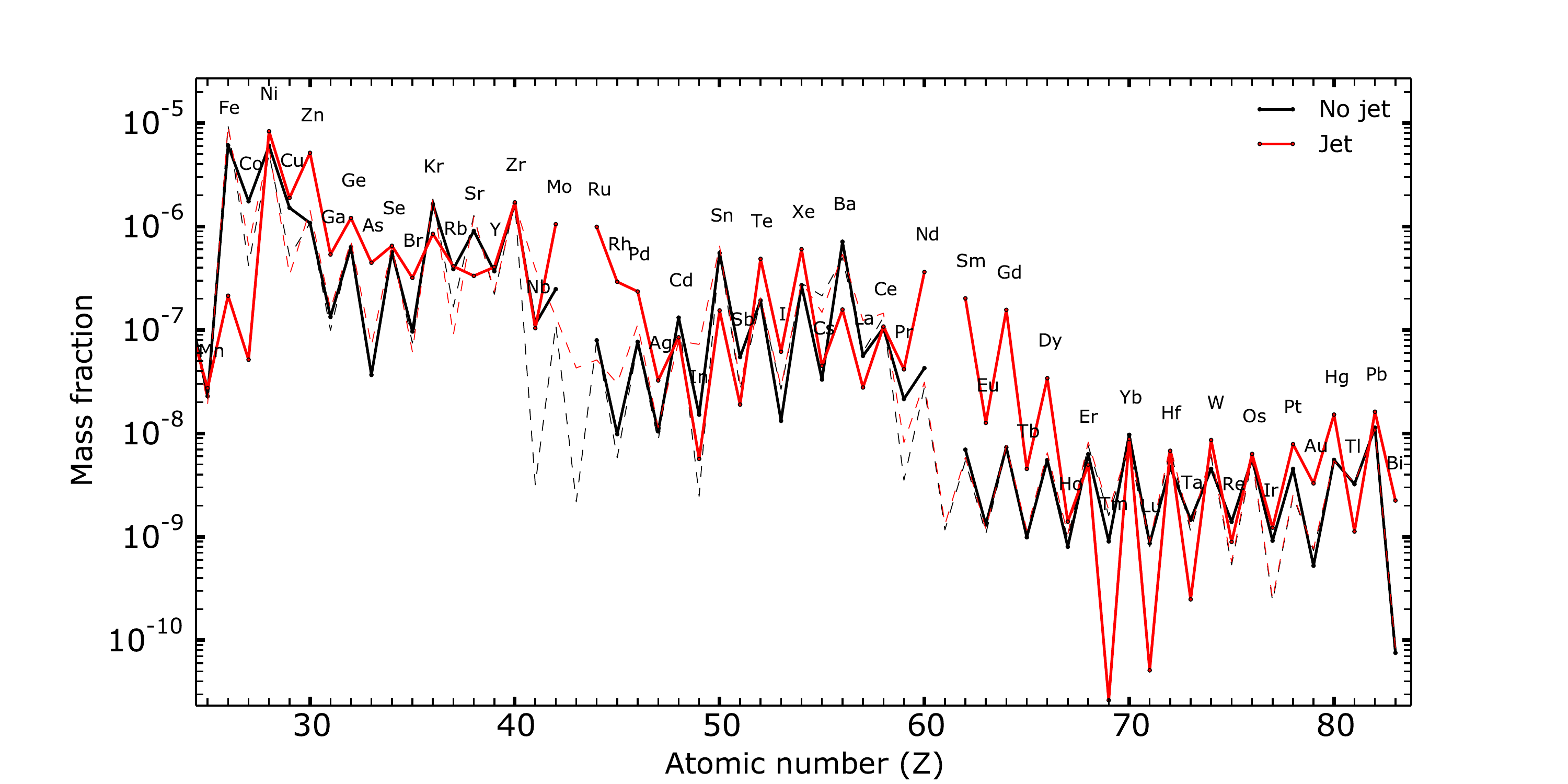}
   \caption{Effect of the jet on heavy elements for the mass particle at $M_{\rm r,ini} = 16.53$~$M_{\odot}$ and $\theta_{\rm ini} = 3.15^{\circ}$ (the same mass particle as in Fig.~\ref{abprof}, \ref{flow1}, \ref{nucchart}, and \ref{aba}). 
The solid red line shows the composition after the jet explosion and after beta decays.
The solid black line shows the composition if no jet explosion occurred (still after beta decays).
Dashed lines show the patterns before beta decays. 
   }
\label{abz}
    \end{figure}

As a test, we ran two spherical explosions: one with the same total energy ($E_{\rm TOT} = 10^{52}$~erg) and one with $E_{\rm TOT} = 10^{53}$~erg. For the $10^{52}$~erg spherical explosion, the bottom of the helium shell is heated up to $\sim 0.6$~GK. In this case, the $({\alpha,n})$ reactions are weaker and the neutron density goes up to $\sim 10^{15}$~cm$^{-3}$. It leads to a much smaller neutron exposure that  slightly affects the distribution of heavy elements. 
For a $10^{53}$~erg spherical explosion, the bottom of the helium shell is heated up to $\sim 1$~GK, which leads to a maximal neutron density of about $\sim 10^{19-20}$~cm$^{-3}$, that is, similar to the jet-explosion discussed throughout this paper. In this case, the abundances of heavy elements are significantly affected in the same way as in Fig.~\ref{aba} and \ref{abz}. 
This means that $\gtrsim 10$ times more energetic spherical explosions could lead to a similar nucleosynthesis as  the $10^{52}$~erg jet-like explosion considered in this work. 
We note that a $10^{53}$~erg explosion corresponds to an extreme case which may not be realistic\footnote{The strongest explosions detected have $10^{52} < E < 10^{53}$~erg 
\citep[see e.g. Figure 2 in][]{nomoto13}.
}. 
Consequently, the nucleosynthetic signature of a boosted n-process (as discussed throughout this paper)
very likely points towards jet-like, or at least aspherical, explosions. 
We plan to explore the effect of various opening angles and explosion energies in a future work.

\section{Signature in the metal-poor star CS29528-028}\label{fitcs}

   \begin{figure}
   \centering
       \includegraphics[scale=0.345, trim = 0cm 0cm 0cm 0cm]{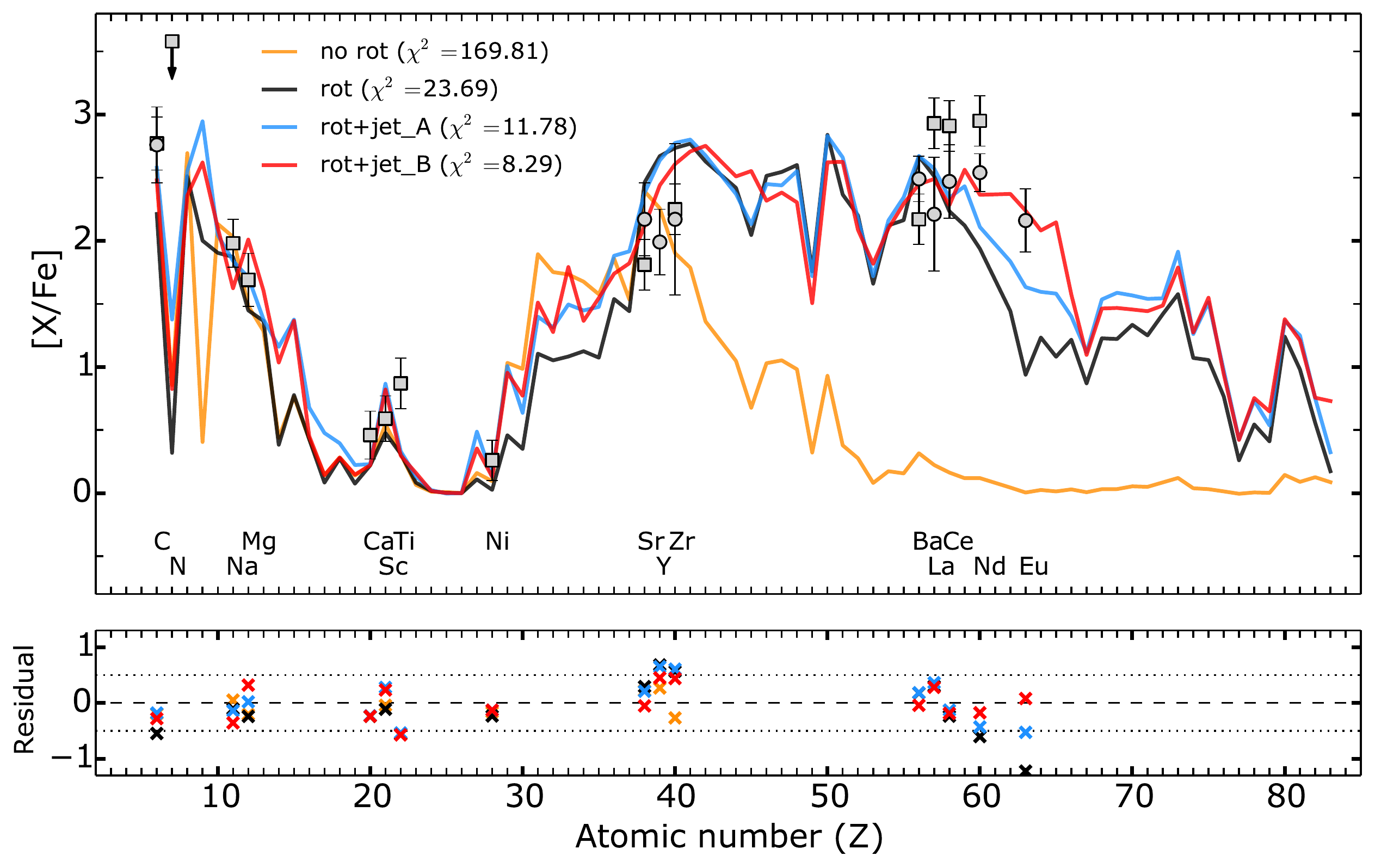}
   \caption{
   Best fits (top panel) and their residuals (bottom panel) for the metal-poor star CS29528-028. The squares are the abundances from \cite{aoki07}, and the circles are from \cite{allen12}. 
   Only an upper limit was derived for nitrogen. 
     When two measurements were available for one element, the most recent one \citep{allen12} was selected to calculate $\chi^2$ and used to plot the residuals. 
The four patterns show the best fits and their $\chi^2$ value for four models (cf. text for details). All models were considered after beta decay. 
  }
\label{cs}
    \end{figure}

The chemical signature of jet-like explosions from rotating stars may be seen at the surface of observed metal-poor stars that have formed with the material ejected by previous sources. 
Here, we focus on CS29528-028 which has [Fe/H]~$-2.12$, $T_{\rm eff} = 7100$~K, and $\log~g = 4.27$ \citep{allen12}. It was classified as a carbon-enhanced metal-poor star with heavy abundances midway between the s- and r-process \citep[CEMP-r/s][]{beers05} because of its high [C/Fe]~$= 2.76$, enrichment in trans-Fe elements, and intermediate [Ba/Eu]~$=0.33$ \citep{aoki07, allen12}. \cite{aoki07} reported a [Na/Fe] ratio of 2.68, but with a suggested NLTE correction of $-0.7$, thus giving [Na/Fe]$_{\rm NLTE} = 1.98$.

\begin{figure}
   \centering
       \includegraphics[scale=0.6, trim = 0cm 0cm 0cm 0cm]{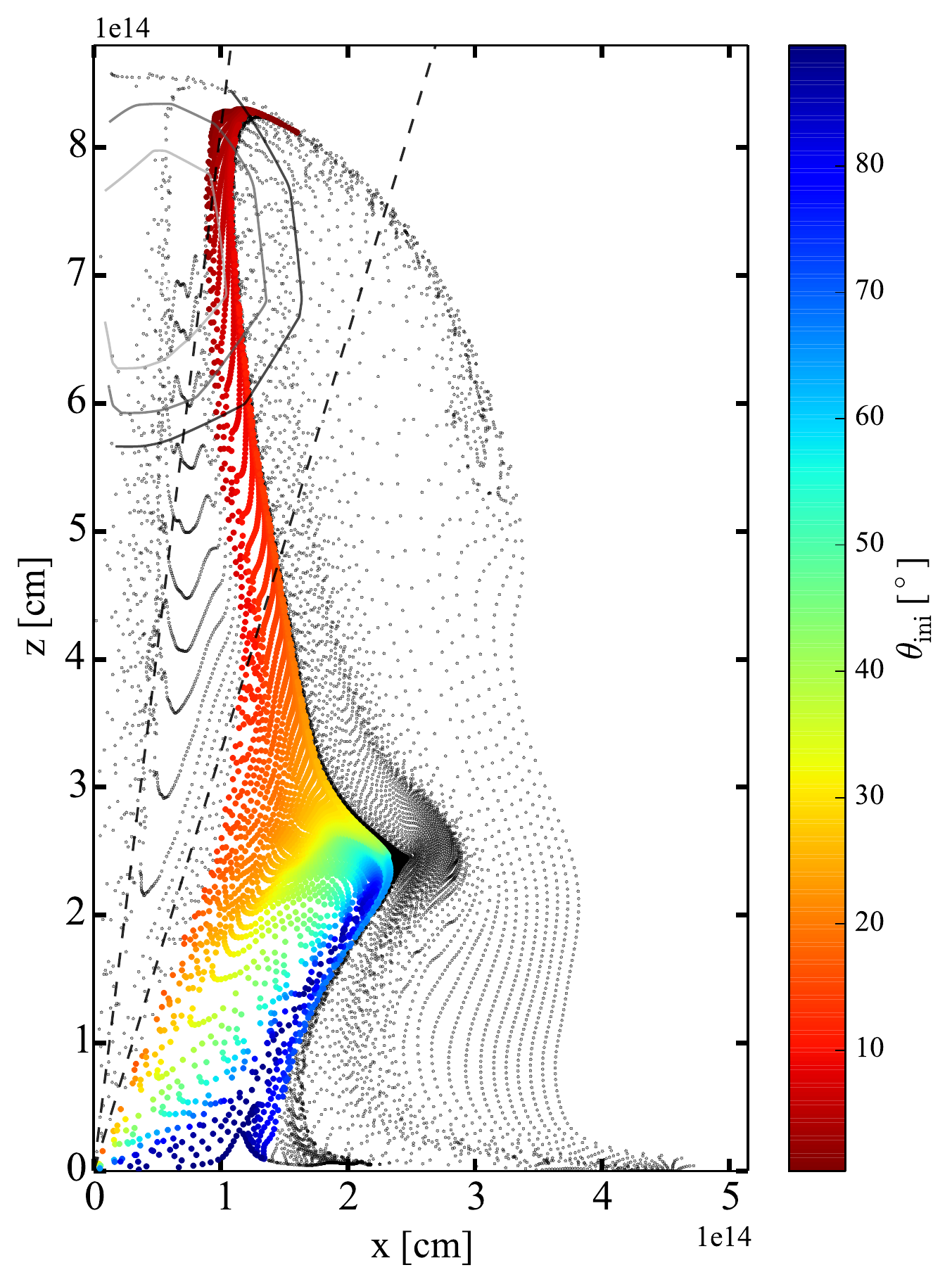}
   \caption{Location of the mass particles at $t = 5 \times 10^{5}$~s. The coloured particles represent the particles originally in the helium shell. The colourmap shows the initial angles of these particles. 
   The two dashed lines show $\theta=7$ and 17~$^{\circ}$. 
   Iso-velocity contours are shown for particles that have the highest velocities ($v_{\rm fin} > 0.04$~c). Lighter contours correspond to higher velocities. 
  }
\label{hefin}
    \end{figure}

We compared the abundances of CS29528-028 with four models. 
The first and second ones are 40~$M_{\odot}$ non-rotating and rotating models, respectively, which did not experience jet-induced nucleosynthesis (orange and black lines in Fig.~\ref{cs}). The mass cut was set at a mass coordinate of 1.93~$M_{\odot}$ (just as the jet model, cf. Sect.~\ref{meth}). 
The ejecta of the non-rotating (rotating) model was mixed with 150 (400) $M_{\odot}$ of ISM material to produce the  fits shown in Fig.~\ref{cs}. 
The third model (rot+jet\_A, blue pattern) shows the integrated chemical composition of all ejected mass particles after the jet.
For the fourth model, we searched for the best fit by considering the dependance of the yields in the final angle $\theta_{\rm fin}$ and final velocity $v_{\rm fin}$ of the ejected material. 
Indeed, a neighbouring halo hosting a future star may not be enriched by the full ejecta, but only by a part of it, having a certain angle range. 
Also, the higher velocity material is more likely to reach and enrich a neighbouring halo. 
The rot+jet\_B model (red pattern in Fig.~\ref{cs}) shows the best fit when the final angle range and the velocity threshold $v_{\rm th}$ are left
as free parameters; it is important to specify that only the mass particles with $v_{\rm fin} > v_{\rm th}$ were considered in this case. 
This model corresponds to the integration of the mass particles with $7<\theta_{\rm fin}<17$~$^{\circ}$ 
and with $v_{\rm fin} >0.04$~c. 
The mass particles in this angle range are located in between the two dashed lines shown in Fig.~\ref{hefin}. 
The mass particles with $v_{\rm fin} >0.04$~c are enclosed by the outermost contour shown in Fig.~\ref{hefin}.

Without rotation, the enrichment in elements with Z~$>$~40 cannot be reproduced. Rotation provides elements from the second s-process peak but lacks Nd and Eu. 
Jet-induced nucleosynthesis boosts the production of the elements beyond the first and second s-process peaks (cf. Sect.~\ref{jetheavy}) and, therefore, can provide the required Nd and Eu (rot+jet\_B model). 
However, if the ejecta is fully mixed (rot+jet\_A model), Nd and Eu are underproduced. 
This is because in this case 
the material processed by the jet is diluted with a large amount of unprocessed (or slightly processed) material, which blurs the nucleosynthesis signature of the jet.

The rot+jet\_B model requires that the fastest ejecta in a given angle range reaches a nearby halo hosting the future CS29528-028 star. 
The requirements of this scenario are stringent, meaning that it may be a rare event.

\section{Discussion}\label{disc}

\subsection{The efficiency of the n-process}

First, the strength of the n-process very much depends on the amount of $^{22}$Ne in the helium-shell at the pre-SN stage.
In a non-rotating model, this amount is directly linked to the initial amount of CNO elements: During the main-sequence, the CNO cycle transforms most of the initial CNO into $^{14}$N and the $^{14}$N is converted into $^{22}$Ne during helium burning by the chain $^{14}$N($\alpha,\gamma$)$^{18}$F($\beta^+$)$^{18}$O($\alpha,\gamma$)$^{22}$Ne. 
Therefore, higher metallicity stars, which contain more CNO at birth, are more likely to experience a more efficient n-process. 
At solar metallicity, the initial sum of CNO elements in a mass fraction is about $10^{-2}$, which leads to a $^{22}$Ne mass fraction at the pre-SN stage of about $10^{-2}$ at maximum. This is similar to the case discussed here (Fig.~\ref{abprof}), meaning that an energetic explosion at solar metallicity should also lead to an efficient n-process.

Rotation-induced mixing synthesises additional primary $^{22}$Ne during the core-helium burning phase (cf. Sec.~\ref{rotind}). Some of it is burnt during the evolution, thus boosting the s-process, and some of it is left at the pre-SN stage. 
If the leftover $^{22}$Ne is burnt during the explosion, that boosts the n-process. 
Rotation can therefore provide a significant amount of $^{22}$Ne at the pre-SN stage, even at low metallicity, and thus leads to a boosted n-process at low metallicity. 

Nevertheless, the n-process scales with the abundances of existing heavy seeds (e.g. Fe, Sr, Ba) in the helium shell at the pre-SN stage. 
This process merely redistributes the abundances of heavy elements in the helium shell. 
If the abundances of the heavy seed are very small in the shell at the pre-SN stage, the n-process continues to operate but the absolute abundances of heavy elements stay very small.

Finally, the strength of the n-process is highly dependent on the peak temperature reached in the helium-shell during the explosion. It the peak temperature is too low, the $^{22}$Ne($\alpha,n$) reaction is not efficiently activated and the n-process stays weak, regardless of the pre-SN $^{22}$Ne abundance. For a given energy, an aspherical explosion allows for a higher peak temperature to be reached in the helium shell (cf. Sect.~\ref{sph}).

For instance, \cite{rauscher02} reported that for non-rotating solar metallicity $\sim 20$~$M_{\odot}$ stars exploding as spherical type II SNe with $10^{51} \lesssim E \lesssim 2 \times 10^{51}$~erg, the n-process only slightly redistributes the heavy elements at the bottom of the helium shell (cf. their section 5.4). These models contain $^{22}$Ne, but it is likely not significantly burnt during the explosion.

\subsection{The n- and the i-process}

In the nuclear chart, the boosted n-process that is studied here populates a region midway between the s-process and r-process regions (Fig.~\ref{nucchart}). However, this process differs from the i-process. 
The latter is associated with hydrogen ingestion in a convective helium burning region, which triggers the chain $^{12}$C($p,\gamma$)$^{13}$N($\beta^{+}$)$^{13}$C($\alpha,n$) and leads to neutron densities in between the s- and r-process \citep[typically $n_n \sim 10^{13}$~cm$^{-3} - 10^{15}$~cm$^{-3}$,][]{cowan77,malaney86,dardelet14,clarkson18,denissenkov19,hampel19}. 
Typical neutron exposures are $1 \lesssim \tau \lesssim 20$~mbarn$^{-1}$ and the timescale involved is $\gtrsim 1$~day. 
By contrast, in the case of the n-process, the neutrons are mainly provided by the $^{22}$Ne($\alpha,n$) reaction; the timescale is shorter ($\sim 1$~s) because of the explosive nature of the event. Also, the neutron density is higher and the neutron exposure is similar or lower (Fig.~\ref{nexp}).

\subsection{On the production of Pb in massive stars}

The abundance of Pb is challenging to determine in metal-poor stars because it generally relies on a single line, which may be blended by a CH line. 
Although Pb is not detected in CS29528-028, some CEMP-r/s stars have a high Pb abundance \citep[e.g.][]{placco13}. 

We recall that any Pb that is synthesised during the massive star evolution in the helium shell should not be affected by the passage of the jet (cf. Sect.~\ref{jetheavy}). Thus, if any Pb is synthesised in massive stars, it should occur during stellar evolution. 
The rotating massive stellar models of \cite{limongi18} and \cite{banerjee19} produce a significant amount of Pb during their lives, while only small or modest amounts are synthesised in similar models from \cite{frischknecht12} , \cite{frischknecht16}, \cite{choplin18}, \cite{choplin20}, and from this work (cf. Fig~\ref{presn}). 
Among the important parameters impacting the production of Pb is the treatment of rotation in the stellar interior, which varies from code to code. 
An efficient internal rotational mixing during stellar evolution favours the production of Pb: It provides more $^{22}$Ne and $^{13}$C (cf. Sect.~\ref{scenar}) and thus increases the neutron source ($^{13}$C, $^{22}$Ne) over seed (e.g. $^{56}$Fe) ratio, thus boosting the production of heavier s-elements, such as Pb. 
Unfortunately, the physics of rotation still suffers from important uncertainties and therefore a firm answer cannot be given now.

\section{Summary and conclusions}\label{concl}

We inspected the combined effect of fast rotation and jet-like explosion on the helium shell nucleosynthesis of a low metallicity 40~$M_{\odot}$ star. 
As has already been found, rotation boosts the s-process during the evolution and leaves a high amount of unprocessed $^{22}$Ne in the helium shell at the pre-supernova stage. 
An energetic jet-like explosion heats the helium shell up to $\sim 1.5$~GK. It efficiently activates ($\alpha,n$) reactions (especially $^{22}$Ne($\alpha,n$)) and leads to neutron densities of $10^{19} - 10^{20}$~cm$^{-3}$ for $\sim 0.1$~second. 
This neutron burst 
shifts the s-process pattern towards heavier elements. 
In particular, the production of Mo, Ru, and Rh (right after the first s-process peak) and Nd, Sm, Eu, and Gb (right after the second s-process peaks) is significantly boosted.
A spherical explosion, with the energy equally distributed to all angles, does not heat the helium shell sufficiently enough to produce these types of high neutron densities unless the explosion energies is extremely high, about $10^{53}$~erg.

Overall we have shown that the helium shell of a rotating massive star experiences two successive efficient neutron capture processes: an efficient s-process during the evolution and an efficient n-process during a jet-like explosion.
It gives a material whose chemical composition is midway between the s- and r-process.
Such a scenario may explain the metal-poor stars showing these types of intermediate abundance patterns. 
In particular, the abundances of the CEMP-r/s star CS29528-028 are compatible with the yields of our jet model, provided a non-homogeneous mixing of the ejecta. 

We also note that the r-process may occur in the jet or accretion disc of jet-like explosions \cite[e.g.][]{winteler12, nishimura15, siegel19}. 
Depending on the degree of mixing between the jet, disc, and stellar mantle material, combinations of s-, n-, and r-process patterns may result and produce a variety of chemical signatures in the interstellar medium.

\begin{acknowledgements} 
The authors thank an anonymous referee for constructive comments that improved the manuscript. 
A.C. acknowledges funding from the Swiss National Science Foundation under grant P2GEP2\_184492. 
B.S.M. acknowledges funding from NASA Grant NNX17AE32G. 
\end{acknowledgements}

\bibliographystyle{aa}
\bibliography{$HOME/Obs/docs_latex/biblio/biblio.bib}

\end{document}